\documentclass[10pt, conference]{IEEEtran}

\usepackage{cite}
\usepackage{amssymb}
\usepackage{amsthm}
\usepackage{amsmath} 			

\usepackage[boxed,linesnumbered]{algorithm2e}
\usepackage{url}
\usepackage{graphicx}
\usepackage{multirow}
\usepackage{tikz}
\usetikzlibrary{arrows,positioning}
\usepackage{rotating}
\usepackage{epstopdf}
\usepackage{pgfplots}
\pgfplotsset{compat=1.13}
\usepackage{subfigure}
\usepackage{fancyvrb}
\usepackage{pifont}
\usepackage[font=small,format=plain,labelfont=bf,up,textfont=it,up]{caption}

\begin{document}

\title{Evaluating data-flow coverage in \\spectrum-based fault localization}

\author{

\IEEEauthorblockN{Henrique L. Ribeiro, Roberto P. A. de Araujo, Marcos L. Chaim}
\IEEEauthorblockA{Software Analysis and Experimentation Group --- SAEG\\
School of Arts, Sciences and Humanities\\
University of S\~ao Paulo\\
\{henriquelemos,roberto.araujo,chaim\}@usp.br}

\and
\IEEEauthorblockN{Higor A. de Souza, Fabio Kon}
\IEEEauthorblockA{Department of Computer Science\\
Institute of Mathematics and Statistics\\
University of S\~ao Paulo\\
\{hamario,kon\}@ime.usp.br}

}

\maketitle

\begin{abstract}
\underline{Background}: Debugging is a key task during the software development cycle. Spectrum-based Fault Localization (SFL) is a promising technique to improve and automate debugging. SFL techniques use control-flow spectra to pinpoint the most suspicious program elements. However, data-flow spectra provide more detailed information about the program execution, which may be useful for fault localization. \underline{Aims}: We evaluate the effectiveness and efficiency of ten SFL ranking metrics using data-flow spectra. 
\underline{Method}: We compare the performance of data- and control-flow spectra for SFL using 163 faults from 5 real-world open source programs, which contain from 468 to 4130 test cases. The data- and control-flow spectra types used in our evaluation are definition-use associations (DUAs) and lines, respectively. 
\underline{Results}: Using data-flow spectra, up to 50\% more faults are ranked in the top-15 positions compared to control-flow spectra. Also, most SFL ranking metrics present better effectiveness using data-flow to inspect up to the top-40 positions. The execution cost of data-flow spectra is higher than control-flow, taking from 22 seconds to less than 9 minutes. Data-flow has an average overhead of 353\% for all programs, while the average overhead for control-flow is of 102\%. 
\underline{Conclusions}: The results suggest that SFL techniques can benefit from using data-flow spectra to classify faults in better positions, which may lead developers to inspect less code to find bugs. The execution cost to gather data-flow is higher compared to control-flow, but it is not prohibitive. Moreover, data-flow spectra also provide information about suspicious variables for fault localization, which may improve the developers' performance using SFL.
\end{abstract}

\begin{IEEEkeywords}
Fault localization; Spectrum-based; Debugging; Structural testing; Data-flow; Control-flow.
\end{IEEEkeywords}

\section{Introduction}

Testing and debugging are among the most expensive tasks during the development cycle 
\cite{tassey2002}. Debugging consists of locating and fixing program bugs (faults). 
These activities are accomplished with the help of static (e.g., 
source code, bug reports) and dynamic information (e.g., variable values, test results). 
Nevertheless, a developer may spend too much time 
trying to understand and  locate a bug, affecting considerably the overall cost
of the software. This is so because fault localization is in general a tedious, time-consuming, 
and error-prone manual task \cite{mao2014,dandan2014}. Several fault localization techniques have been 
proposed to improve the developer's productivity during the debugging process \cite{zeller2009}.

Spectrum-based fault localization (SFL) is a promising debugging technique due to its relative low overhead in test execution time. SFL techniques use data collected 
during a test suite execution, called program spectra, to infer which source code elements are more likely to contain a bug \cite{jones2002,santelices2009,mao2014}. 
Program spectra\footnote{Henceforth, we use indistinctly the terms  \textit{spectra} 
and \textit{coverage}.} comprise program elements, from fine-grained to coarse ones, that are tracked during a test case execution. They include statements, blocks, predicates, definition-use associations (DUAs), and function or method calls that are associated with each test case.
SFL is based on strategies to pinpoint faulty program elements. Most of them make use of  metrics \cite{abreu2007,naish2009,jones2002,wong2010,gonzalez-sanchez2007} to rank suspicious elements. The rationale is that program elements frequently executed in failing test cases are more likely to be faulty. Thus, the frequency with which elements are executed in passing and failing test cases is used to calculate a suspiciousness score for each element. Subsequently, these suspicious elements are mapped onto lines of code likely to be faulty.

Thus, SFL is built upon two foundations: spectra and ranking metrics. Many studies have been conducted to determine which is the best metric to support SFL \cite{pearson2017,ma2014}. Ochiai \cite{abreu2007} has been suggested as the most effective metric for fault localization \cite{ma2014}. However, a very recent study shows that the best metrics are indistinctive for real faults, meaning that there is hardly any difference in which metric one chooses since the results will be very similar \cite{pearson2017}. 

It remains then to tackle the role of the spectra in SFL. Few studies compared the use of different spectra in fault localization. Santelices et al. \cite{santelices2009} suggest that DUA spectrum performs better than statements and branches to locate bugs. Masri shows that DIFA (\textit{Dynamic Information Flow Analysis}) spectrum, along with DUA and branch, performed better than statement spectrum \cite{masri2010}. DIFA consists of interactions occurred in an execution, and comprises data and control dependencies from statements and variables. Both DUAs and DIFAs are data-flow spectra.

These  studies, though, used  only small- to medium-sized programs and artificial faults in their assessments. In Santelices et al. \cite{santelices2009}, most of the programs belongs to the Siemens Suite \cite{hutchins1994}, which is composed of programs with few hundred lines of code (LOC); in Masri \cite{masri2010}, two programs with less than 10 KLOC were used in the evaluation. Both studies only utilize artificial faults. Pearson et al. \cite{pearson2017}, however, show that many claims verified with artificial faults do not hold for real bugs.

There is a reason why most SFL studies \cite{renieris2003,jones2007,abreu2007,chilimbi2009} use only control-flow spectra, namely, statement, predicate, and branch coverage: cost. To obtain the spectra for fault localization, the original source or object code needs to be modified (instrumented) to track the elements executed by each test case. Moreover, test case results (i.e., fail or success) should also be recorded.
The cost to obtain data-flow spectra is higher because definition-use associations (DUAs), for example, comprise all value assignments (variable definition) and their subsequent references (variable uses) in a program  \cite{hutchins1994,santelices2009}. As a result, there are much more DUAs to track at run-time than statements, which implies both execution and memory overheads during test case execution.
Previous studies show that statements and branches can be monitored with a run-time overhead of 9\% to 18\% while DUAs have a run-time overhead of 66\%-127\% \cite{misurda2005,santelices2007}. Recent results, though, show that DUA spectra can be obtained with fairly moderate execution and memory overheads \cite{chaim2013,araujo2014}. They allow the use of data-flow spectra in fault localization for large and long running programs.

This paper assesses the use of data-flow spectra in fault localization. Because programs have more DUAs than statements or branches, the likelihood of correlating critical DUAs with failing test runs should be higher. Thus, our hypothesis is that data-flow spectra is more effective than control-flow spectra for fault localization.  Moreover, we investigate whether the possible increase on the effectiveness is worth the data-flow extra cost for large programs. We compared data- and control-flow spectra using definition-use associations (DUAs) and statement (line\footnote{The term \textit{line} in this work refers to an executable line or statement.}), respectively. DUAs and lines are two of the least expensive program elements that represent both spectra.

We used the Jaguar SFL tool (\emph{JAva coveraGe faUlt locAlization Ranking})\footnote{github.com/saeg/jaguar}  \cite{ribeiro2018} to compare data- and control-flow spectra in fault localization. Jaguar provides ranking lists for several ranking metrics. It uses JaCoCo\footnote{eclemma.org/jacoco} and BA-DUA\footnote{github.com/saeg/ba-dua} to collect line and DUA spectra, respectively. JaCoCo has been utilized to assess control coverage of large systems. Moir reports that JaCoCo was run against 37,000 Eclipse JDT core tests with an execution overhead of 2\% \cite{moir2011}. BA-DUA, in turn, was able to collect DUA spectra with run-time overhead of 38\% on average for large programs \cite{araujo2014}. 

Differently from previous work, we assess both techniques using 163 real faults from five real open-source programs. The program sizes vary from 10 to 96 KLOC, which contains from 468 to 4130 test cases. Additionally, we compare data- and control-flow spectra using ten SFL ranking metrics. The goal is to assess which spectrum is more effective; i.e., the one that locates more bugs by inspecting fewer lines of code. Furthermore, we compare the efficiency (i.e., run-time overhead) of SFL based on data- and control-flow spectra. The following research questions summarize the issues addressed in this paper:

\begin{enumerate}
\item Which spectrum is more effective to locate bugs: data- or control-flow?
\item Which spectrum is more efficient to locate bugs: data- or control-flow?
\end{enumerate}

Our results indicate that data-flow spectrum ranks more faults in upper positions compared to control-flow spectrum with statistical significance for all ten ranking metrics. 
DUA outperforms line spectrum, classifying more faults in the range of 7 to 40 most suspicious lines in the SFL ranking lists. By investigating from the top 9 to the top 25 most suspicious lines, a practitioner can reach from one quarter to half of all faults using DUA spectrum and Ochiai. To achieve the same result with line spectrum, one needs to inspect 14 more lines. After the 70th position, control-flow surpass data-flow for most ranking metrics. Since each DUA contains from 2 to 3 lines, its performance worsens for faults that are far from the top positions. 

Data-flow cost is high, but it is not prohibitive. It took from 22 seconds to 8:35 minutes to obtain the suspicious lines. However, control-flow spectrum can also be expensive, up to 4:25 minutes. The average time for all programs using data-flow and control-flow are 2:58 and 1:18 minutes, respectively. Data-flow has an average overhead of 353\% for all programs, while the average overhead for control-flow is of 102\%. 

The results suggest that data-flow spectrum can leverage SFL; however, practitioners should be educated to inspect more code---from 10 to 25 suspicious lines---to take advantage of SFL.

The remainder of the paper is organized as follows. Section~\ref{back} presents concepts and definitions of control- and data-flow spectra, and Spectrum-based Fault Localization (SFL). Sections~\ref{exp}, \ref{res}, \ref{dis}, and \ref{threats} describe, respectively, the experiment design, the results, the discussion, and the threats to the validity of our experimental assessment. The related work is discussed in Section~\ref{related}. We draw our conclusions and present future work in Section~\ref{concl}.

\section{Background}
\label{back}
In what follows, we present an example to illustrate the concepts of control- and 
data-flow spectra, and spectrum-based fault localization. 

\subsection{Running example}

Figure \ref{code_example} shows the code of a simple function, named \textit{max} \cite{chaim2013}. 
The first three columns represent line, statement, and block 
numbers, respectively. The function receives two parameters: the first is an \texttt{int array} and 
the second is the array size. It should return the largest number of an array, 
but there is a fault at line 4: \texttt{array[++i]} should be \texttt{array[i++]}; that is, the 
increment (\texttt{++}) must come \emph{after} the variable \texttt{i}. This causes 
variable \texttt{max} to be assigned to the second position of the array.

\begin{figure}[ht]
\centering \small
\begin{tabular}{c|c|c|l}
  \hline 
  Line & Statement & Block & Code \\
  \hline 
  1 & - & - & int max(int[] array, int length) \\
  2 & - & 1 & \{ \\
  3 & 1 & 1 & \hspace{0.2 cm} int i = 0; \\
  4 & 2 & 1 & \hspace{0.2 cm} int max = array[++i]; \textbf{//array[i++];} \\
  5 & 3 & 2 & \hspace{0.2 cm} while(i $<$ length) \\
  6 & - & 3 & \hspace{0.2 cm} \{ \\
  7 & 4 & 3 & \hspace{0.6 cm} if(array[i] $>$ max) \\
  8 & 5 & 4 & \hspace{0.9 cm} max = array[i]; \\
  9 & 6 & 5 & \hspace{0.6 cm} i++; \\
  10 & - & 5 & \hspace{0.2 cm} \} \\
  11 & 7 & 6 & \hspace{0.2 cm} return max; \\
  12 & - & 6 & \} \\
  \hline
\end{tabular}
\caption{Code of \textit{max} program\label{code_example}}
\end{figure}

\subsection{Control-flow spectra}
\label{sec:control-flow}

Fault localization techniques use different types of control-flow spectra: statements are 
executable lines of code; basic blocks (or simply blocks) are sets of statements that 
are always executed together; branches are statements that transfer the control-flow 
execution among blocks. 


Control-flow information of a program is represented by a \emph{graph} with \emph{nodes} and 
\emph{edges}, in which a \emph{node} represents a \emph{basic block} and an \emph{edge} represents 
a \emph{branch}. Figure \ref{max-cfg} shows the control-flow graph of the \emph{max} program.

Table~\ref{tab:cov-criteria} describes, in the \emph{all-nodes} and \emph{all-edges} columns, the 
control-flow spectra that can be tracked during the execution of \emph{max}'s test suite. 
The nodes and edges are also known as the testing requirements of structural testing criteria 
all-nodes and all-edges \cite{rapps1985}. A node (edge) is considered as covered if there is a 
test case that traverses a path that includes such a node (edge). All-nodes and all-edges 
criteria require that every node and every edge of a program, respectively, must be covered by at 
least one test case to be satisfied.

\subsection{Data-flow spectra}
\label{sec:data-flow}

\begin{figure}[ht]
\centering
\scalebox{1.0}{
\begin{tikzpicture}[
    ->,
    shorten >=2pt,
    >=stealth,
    node distance=1cm,
    noname/.style={
      circle,
      minimum width=2em,
      minimum height=2em,
      draw,
	line width = 1.0pt
    },
	labelnode/.style={
		rectangle,
		draw=none,
		text width={width("aaaafor techniques")+2pt},
		font=\small 
	}
]
\node[noname] (1)                                              {1};
\node[noname] (2) [node distance=0.75cm, below=of 1]              {2};
\node[noname] (3) [node distance=1cm and 5mm,below left=of 2]  {3};
\node[noname] (4) [node distance=0.7cm and 5mm,below left=of 3]{4};
\node[noname] (5) [node distance=1.5cm,below =of 3]            {5};
\node[noname] (6) [node distance=1cm and 5mm,below right=of 2] {6};
\node[labelnode] (def1) [node distance=0.35cm, left =of 1] {def=\{i,array,length,max\}};
\node[labelnode] (cuse0) [node distance=-0.05cm, right =of 1] {c-use=\{i,array\}};
\node[labelnode] (puse1) [node distance=-0.6cm, yshift=-0.6cm, left =of 2] {p-use=\{i,length\}};
\node[labelnode] (puse2) [node distance=0.05cm, yshift=-0.6cm, right =of 2] {p-use=\{i,length\}};
\node[labelnode] (puse3) [node distance=-0.2cm, yshift=-0.4cm, left =of 3] {p-use=\{i,array,max\}};
\node[labelnode] (def2) [node distance=-1.4cm, yshift=0.15cm, left =of 4] {def=\{max\}};
\node[labelnode] (cuse1) [node distance=-0.9cm, yshift=-0.25cm, left =of 4] {c-use=\{i,array\}};
\node[labelnode] (cuse2) [node distance=-0.05cm, right =of 6] {c-use=\{max\}};
\node[labelnode] (puse4) [node distance=0.25cm, xshift=1.45cm, below =of 3] {p-use=\{i,array,max\}};
\node[labelnode] (def3) [node distance=0.01cm, xshift=0.9cm, below =of 5] {def=\{i\}};
\node[labelnode] (cuse3) [node distance=0.35cm, xshift=0.8cm, below =of 5] {c-use=\{i\}};
\path (1) edge                   node {} (2)
      (2) edge                   node {} (3)
      (2) edge                   node {} (6)
      (3) edge                   node {} (4)
      (3) edge                   node {} (5)
      (4) edge                   node {} (5);
\draw[->] (5) to[out=350,in=10,looseness=3.0] (2);
\end{tikzpicture}
}
\caption[Control-flow graph of \emph{max} program]{Control-flow graph of \emph{max} program\label{max-cfg}}
\end{figure}
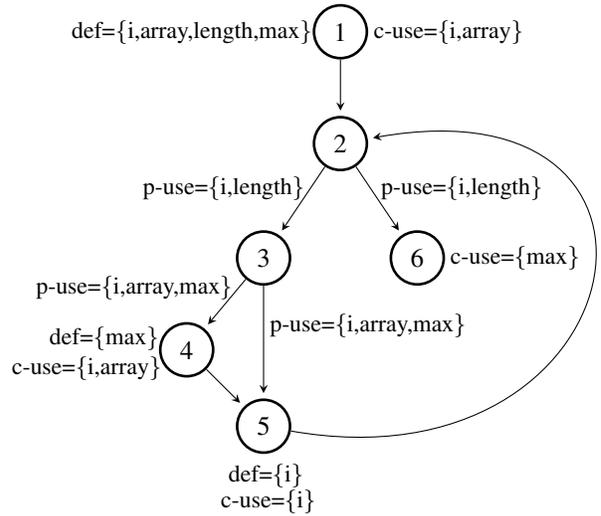

Data-flow spectrum concerns paths in a program between every point in which a value is assigned to a 
variable and its subsequent references. When a variable receives a new value, it is said that 
a \textit{definition} has occurred; a \textit{use} of a variable happens when its value is 
referred to. A distinction is made between a variable referred to compute a value and to 
compute a predicate. When referred to in a predicate computation, it is called a \textit{p-use} 
and is associated with edges; otherwise, it is called a \textit{c-use} when associated with 
nodes. A \textit{definition-clear} path with respect to (wrt) a variable $X$ is a path where 
$X$ is not redefined in any node in the path, except possibly in the first and last ones. 
Figure~\ref{max-cfg} contains \emph{max}'s control-flow graph annotated with data-flow
information.

Data-flow spectra can be formally defined using the concept of \textit{definition-use 
associations} (DUAs) \cite{rapps1985}. The triple $D$ = ($d$, $u$, $X$), called c-use DUA, represents a 
data-flow testing requirement involving a definition in node $d$ and a c-use in node $u$ 
of variable $X$ such that there is a definition-clear path wrt $X$ from $d$ to $u$. 
Likewise, the triple $D$ = ($d$, ($u'$, $u$), $X$), called  p-use DUA, represents the 
association between a definition and a p-use of a variable $X$. In this case, a 
definition-clear path ($d$,$\ldots$,$u'$,$u$) wrt $X$ should exist.

The data-flow spectrum is also the testing requirements of the  \emph{all-uses}  
testing criterion \cite{rapps1985}. This criterion requires the set of paths executed by the 
test cases of a test set $T$ to include a definition-clear path for each DUA ($d$, $u$, $X$) 
or ($d$, ($u'$, $u$), $X$) of a program $P$. The DUAs exercised by a test set are said to be 
\textit{covered} by it. Table~\ref{tab:cov-criteria} contains the DUAs required for the 
\emph{max} program.

\begin{table*}[ht]
\centering 
\caption{Data- and control-flow spectra for the max program}
\label{tab:cov-criteria}
\begin{tabular}{c|c|ccc}
	\hline 
	All-nodes & All-edges & \multicolumn{3}{|c}{All-uses} \\
	\hline 
	1 & (1,2) & (1, 6, max) &(4, 6, max) &(1, (3,4), max)\\
	2 & (2,3) & (1, (3,5), max) &   (4, (3,4), max) &  (4, (3,5), max)\\
	3 & (2,6) & (1, 1, i) & (1, 4, i) & (1, 5, i) \\
	4 & (3,4) & (1, (2,3), i) & (1, (2,6), i) & (1, (3,4), i) \\
	5 & (3,5) & (1, (3,5), i) & (5, 4, i) & (5, 5, i) \\
	6 & (4,5) & (5, (2,3), i) & (5, (2,6), i) & (5, (3,4), i) \\
	  & (5,2) & (5, (3,5), i) & (1, 1, array) & (1, 4, array) \\
	  &       & (1, (3,4), array) & (1, (3,5), array) & (1, (2,3), length) \\
	  &       & (1, (2,6), length) &  & \\
	\hline
\end{tabular}
\end{table*}

\subsection{Spectrum-based fault localization}
\label{sec:sfl}

Program spectrum or coverage \cite{reps1997,abreu2007} consists of program elements (e.g., 
statements, blocks, predicates, DUAs, methods) covered during a test case execution 
\cite{elbaum2001,harrold1998}. Spectrum-based fault localization techniques use ranking 
metrics to assign suspiciousness scores to these elements. The outcome is a list of elements 
ranked in descending order of suspiciousness. 

Several ranking metrics have been proposed or used to indicate elements most likely to contain faults 
\cite{jones2002,abreu2007,debroy2013}. The most effective metrics are characterized by 
identifying suspicious code excerpts from the frequency of elements exercised in passing 
and failing test cases. 
The ranking metrics use four parameters to calculate the suspiciousness of an executed program 
element. These parameters represent the four possible states during a test execution: 
$c_{ef}$ indicates the number of times a program element ($c$) is executed ($e$) in 
failing ($f$) test cases, $c_{nf}$ is the number of times an element is not ($n$) executed 
by failing ($f$) test cases, $c_{ep}$ is the number of times an element is executed ($e$) 
by a passing ($p$) test case, and $c_{np}$ represents the number of times an element is 
not ($n$) executed by passing ($p$) test cases.

\begin{table*}
\centering
\caption{Ranking metrics for fault localization\label{tb:metrics}}{
\begin{tabular}{cc|cc}
  \hline 
  Ranking metric & Formula & Ranking metric & Formula \\
  \hline 
  Ochiai\cite{abreu2007} & \scalebox{1.25}{$\frac{c_{ef}}{\sqrt{(c_{ef} + c_{nf})(c_{ef} + c_{ep})}}$} & Jaccard\cite{abreu2007} & \scalebox{1.25}{$\frac{c_{ef}}{c_{ef} + c_{nf} + c_{ep}}$} \\
  Kulczynski2\cite{naish2009} & \scalebox{1.25}{$\frac{1}{2}\left ( \frac{c_{ef}}{c_{ef} + c_{nf}} + \frac{c_{ef}}{c_{ef} + c_{ep}} \right )$} & Zoltar\cite{gonzalez-sanchez2007} & \scalebox{1.25}{$\frac{c_{ef}}{c_{ef} + c_{nf} + c_{ep} + 10000 \cdot \frac{c_{nf} c_{ep}}{c_{ef}}}$}\\
  McCon\cite{naish2009} & \scalebox{1.25}{$\frac{c_{ef}^2 - c_{nf}c_{ep}}{(c_{ef} + c_{nf})(c_{ef} + c_{ep})}$} & Minus\cite{xu2011} & \scalebox{1.25}{$\frac{\frac{c_{ef}}{c_{ef} + c_{nf}}}{\frac{c_{ef}}{c_{ef} + c_{nf}} + \frac{c_{ep}}{c_{ep} + c_{np}}} - \frac{1 - \frac{c_{ef}}{c_{ef} + c_{nf}}}{1 - \frac{c_{ef}}{c_{ef} + c_{nf}} + 1 - \frac{c_{ep}}{c_{ep} + c_{np}}}$} \\
  $O^p$\cite{naish2009} & \scalebox{1.25}{$c_{ef} - \frac{c_{ep}}{c_{ep} + c_{np} + 1}$} & DRT\cite{chaim2003} & \scalebox{1.25}{$\frac{c_{ef}}{1 + \frac{c_{ep}}{\mid T \mid}}$}, where $\mid T \mid$ is the size of test suite \emph{T}\\
  Tarantula\cite{jones2002} & \scalebox{1.25}{$\frac{\frac{c_{ef}}{c_{ef} + c_{nf}}}{\frac{c_{ef}}{c_{ef} + c_{nf}} + \frac{c_{ep}}{c_{ep} + c_{np}}}$} & Wong3\cite{wong2010} & \scalebox{1.25}{$c_{ef} - p$}, where $p = \begin{cases} c_{ep} & \text{ if } c_{ep} \leq 2 \\ 2 + 0.1(c_{ep} - 2) & \text{ if } 2 < c_{ep} \leq 10\\ 2.8  + 0.001(c_{ep} - 10) & \text{ if } c_{ep} > 10 \end{cases}$ \\
  \hline
\end{tabular}}
\end{table*}

SFL ranking metrics have been studied by different authors, 16 of them are listed by Mao et al. 
\cite{mao2014}. In Table~\ref{tb:metrics}, we present the ten SFL ranking metrics utilized in our 
empirical assessment. Table~\ref{tab:testcases} shows five test cases ($t_1, t_2$, $t_3$, $t_4$, 
and $t_5$) of the \textit{max} program. The test cases $t_4$ and $t_5$ fail, while $t_1, t_2$, 
and $t_3$ pass. Table~\ref{tab:cov-ochiai} presents the statement and the block spectra 
(indicated by the bullets associated with each statement, block, and test case) of \textit{max}, 
the number of times whether an element is executed or not by failing and passing test cases, 
and the suspiciousness score assigned using Ochiai in the $M_O$ column. 

The suspiciousness score for block 1 is 0.63, and 0.35 or 0 for the remaining blocks (shown in Table~\ref{tab:cov-ochiai}). For 
Ochiai, the most suspicious elements are  lines 3 and 4 (or statements 1 and 2). In this example, 
control-flow spectra, in particular, blocks, were ranked. These blocks are then mapped onto lines 
3 and 4 to be inspected by the developer. Thus, the top suspicious lines contain the bug. 
Similarly, one could utilize Ochiai to rank DUAs, which can then be mapped onto a list of 
suspicious lines. We describe how we mapped DUAs onto lines in Section~\ref{exp}.

\begin{table}[ht]
\centering
\caption{Test Suite}
\label{tab:testcases}
\begin{tabular}{c|l|c|c}
    \hline 
    $T_n$ & Test & Expected Result & Actual Result\\
    \hline 
    $t_1$ & max( [1,2,3] , 3 ) & 3 & 3 \\
    $t_2$ & max( [5,5,6] , 3 ) & 6 & 6 \\
    $t_3$ & max( [2,1,10] , 3 ) & 10 & 10 \\
    $t_4$ & max( [4,3,2] , 3 ) & 4 & 3 \\
    $t_5$ & max( [4] , 1 ) & 4 & error\\
    \hline
\end{tabular}
\end{table}

\begin{table*}[t]
\centering 
\caption{Program spectra and suspiciousness of the \textit{max} program with Ochiai metric}
\label{tab:cov-ochiai}
\begin{tabular}{c|c|c|ccccc|cccc|c}
  \hline 
  Line & Statement & Block & $t_1$ & $t_2$ & $t_3$ & $t_4$ & $t_5$ & $c_{np}$ & $c_{ep}$ & $c_{nf}$ & $c_{ef}$ & $M_O$ \\
  \hline 
  1 & - & - & $\bullet$ & $\bullet$ & $\bullet$ & $\bullet$ & $\bullet$ & 0 & 3 & 0 & 2 & 0.63\\
  2 & - & 1 & $\bullet$ & $\bullet$ & $\bullet$ & $\bullet$ & $\bullet$ & 0 & 3 & 0 & 2 & 0.63 \\
  3 & 1 & 1 & $\bullet$ & $\bullet$ & $\bullet$ & $\bullet$ & $\bullet$ & 0 & 3 & 0 & 2 & 0.63 \\
  4 & 2 & 1 & $\bullet$ & $\bullet$ & $\bullet$ & $\bullet$ & $\bullet$ & 0 & 3 & 0 & 2 & 0.63 \\
  5 & 3 & 2 & $\bullet$ & $\bullet$ & $\bullet$ & $\bullet$ &  		& 0 & 3 & 1 & 1 & 0.35 \\
  6 & - & 3 & $\bullet$ & $\bullet$ & $\bullet$ & $\bullet$ &  		& 0 & 3 & 1 & 1 & 0.35\\
  7 & 4 & 3 & $\bullet$ & $\bullet$ & $\bullet$ & $\bullet$ &  		& 0 & 3 & 1 & 1 & 0.35  \\
  8 & 5 & 4 & $\bullet$ & $\bullet$ & $\bullet$ & 		&  		& 0 & 3 & 2 & 0 & 0.00  \\
  9 & 6 & 5 & $\bullet$ & $\bullet$ & $\bullet$ & $\bullet$ &  		& 0 & 3 & 1 & 1 & 0.35 \\
  10 & - & 5 & $\bullet$ & $\bullet$ & $\bullet$ & $\bullet$ &  		& 0 & 3 & 1 & 1 & 0.35 \\
  11 & 7 & 6 & $\bullet$ & $\bullet$ & $\bullet$ & $\bullet$ &  		& 0 & 3 & 1 & 1 & 0.35 \\
  12 & - & 6 & $\bullet$ & $\bullet$ & $\bullet$ & $\bullet$ &  		& 0 & 3 & 1 & 1 & 0.35\\
  \hline
  & & & \color{green!70!black}\ding{51} & \color{green!70!black}\ding{51} & \color{green!70!black}\ding{51} & \color{red}\ding{55} & \color{red}\ding{55} & & & & & \\
  \hline
\end{tabular}
\end{table*}

\section{Experiment design}
\label{exp}

We conducted an experiment\footnote{The datasets and results are available at \url{doi.org/10.5281/zenodo.3258116}} to compare the use of data- and control-flow spectra in fault localization. We applied ten ranking metrics to assess the effectiveness of both spectra. In what follows, we describe the programs and bugs used in the experiment, the data collection procedure, and the analysis criteria used to answer our research questions.

\subsection{Subject programs and faults}

We used five real-world programs for the experiment. JFreeChart, Commons Lang, Commons Math, and Joda-Time 
were obtained from the Defects4J database\footnote{\url{github.com/rjust/defects4j}} \cite{just2014}. 
The jsoup's faults were obtained by us from its source code repository\footnote{\url{github.com/jhy/jsoup}} by looking at the commit history. Therefore, all programs used in our experiment are composed of real faults along with their original test cases.

Table~\ref{tab:programs} lists our subject programs, their number of lines of code (KLOC), 
the number of test cases (TC), and the number of faulty versions (Faults). Each version contains a single fault that may be spread in several lines. That is, changes in more than one line were required to fix the bug. In this scenario, each line of a multiple-line fault is deemed as the bug. The rationale is that, if a developer reaches at least one of these lines, s/he locates the bug. 

\begin{table}[ht]
\centering 
\caption{Programs characteristics and number of faults}
\label{tab:programs}
\begin{tabular}{lrrr}
  \hline 
  Program 	& KLOC & TC     & Faults\\
  \hline 
  Commons Lang	& 22   & 2,245  & 30\\
  Commons Math	& 85   & 3,602  & 43\\
  JFreeChart	& 96   & 2,205  & 26\\
  Joda-Time	& 28   & 4,130  & 26\\
  jsoup		& 10   & 468    & 38\\
  \hline 
  Total		& ---  & ---    & 163\\
  \hline
\end{tabular}
\end{table}

However, in selecting faults for our experiment, we only selected bugs for which data-flow spectra was complete. BA-DUA is unable to collect reliable data-flow spectra in two cases: 1) the fault throws an exception and it is located in a code block with a non-handled exception; and 2) the bug is in a single-block method. The first case happens because BA-DUA marks the exercised DUAs whenever the method is exited; in this case, the method is exited in a non-predictable way so that the BA-DUA library does not mark the exercised DUAs as covered. The second case occurs because BA-DUA does not cover DUAs when the definition and use are in the same block. Hence, we removed faults with incomplete data-flow spectra for a fair comparison between spectra.

\subsection{Data collection procedure}

We used Jaguar \cite{ribeiro2018} to generate the lists of suspicious DUAs and lines for each fault. Jaguar is an open source SFL tool that utilizes the instrumentation libraries of two testing tools: BA-DUA \cite{araujo2014} to gather data-flow spectra and JaCoCo to gather control-flow spectra. 
It then calculates the suspiciousness value of each element according to a chosen ranking metric. The cost of determining the suspiciousness values is very low in comparison to the time for test suite execution and spectra data collection. 

We applied this procedure for all ten ranking metrics (see Table~\ref{tb:metrics}) used in this evaluation: Ochiai, Kulczynski2, McCon, $O^{p}$, Tarantula, Jaccard, Zoltar, Minus, DRT, and Wong3. In this work, we did not intend to compare the ranking metrics; thus, we were not exhaustive in the selection of ranking metrics. Instead, we selected  different metrics only to assess whether the spectra's performance is affected by the chosen metric.

We identified the faulty lines of jsoup and the programs from Defects4J by comparing the buggy and fix commits in their repositories. For faults occurred due to missing code, we deemed the previous line of code as the bug site. However, if the previous line contains non-executable code (e.g., comments, blank lines, closing brackets), the line after the change was deemed as the faulty one.
When ties occur, the worst case is considered. It means that if two or more lines have the 
same suspiciousness score, the position of the faulty line includes all tied lines, since 
they have equal chances of being inspected.

As the data-flow spectra are DUAs, we mapped them onto lines to determine their faulty sites. A DUA has always a \textit{definition} line and a \textit{use} line, and possibly a \textit{source} line (for p-use DUAs). To check whether the bug was found, the faulty line is compared to these three lines (definition, use, and source). When the result is false, all these lines are added to the number of lines inspected until finding the fault.

We created a script to check the position of the faults over each respective DUA and line suspiciousness lists generated by Jaguar. It records the number of lines until reaching the faulty line considering the above criteria.

\subsection{Measurement criteria}

To answer our research questions, we measured the effectiveness of data- and control-flow to locate the faults by the \emph{absolute number} of lines inspected until reaching the faults. We restricted the number of lines inspected to 99 lines of code. If a fault is beyond the 99th position we considered that it was not found. Indeed, this is a very large limit since developers in practice inspect only the first picks of the suspiciousness lists \cite{parnin2011,kochhar2016}. 

We also measured the efficiency of data- and control-flow spectra by collecting the \emph{time spent} to create the suspiciousness lists of each faulty version. The total overhead cost comprehends running all unit tests, storing the spectra data, calculating the suspiciousness of each element, and generating the suspiciousness list.

\subsection{Statistical Analysis}

We carried out a statistical analysis to evaluate the results obtained by data- and control-flow spectra with different ranking metrics. First, we applied the Anderson-Darling normality test \cite{anderson1954} to our data, which follows a non-normal distribution. Then, we used the Wilcoxon signed-rank test \cite{wilcoxon1945} for the hypothesis test and Cliff's delta \cite{cliff1993} to measure the effect size. These statistical methods are suitable to use with non-parametric data. The significance level applied was 5\%. The Wilcoxon signed-rank test assesses differences in the magnitude of measurements between the compared techniques. The Cliff's delta calculates the amount of difference between the two techniques. We explain our null and alternative hypotheses for each test in Section~\ref{res}.

\section{Results}
\label{res}

First, we present the effectiveness of data- and control-flow spectra for different ranking metrics and then the run-time overhead to assess the efficiency of the two spectra.

\subsection{Effectiveness of data- and control-flow spectra}

\begin{figure*}
\centering
\mbox{\subfigure[Ochiai]{\scalebox{.35}{\includegraphics{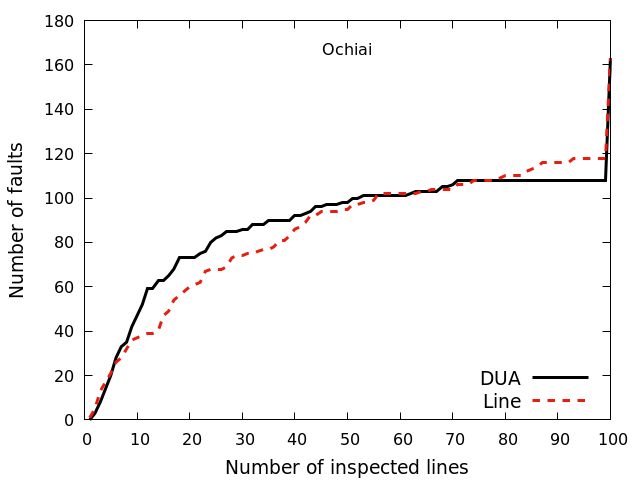}}}\quad
      \subfigure[Kulczynski2]{\scalebox{.35}{\includegraphics{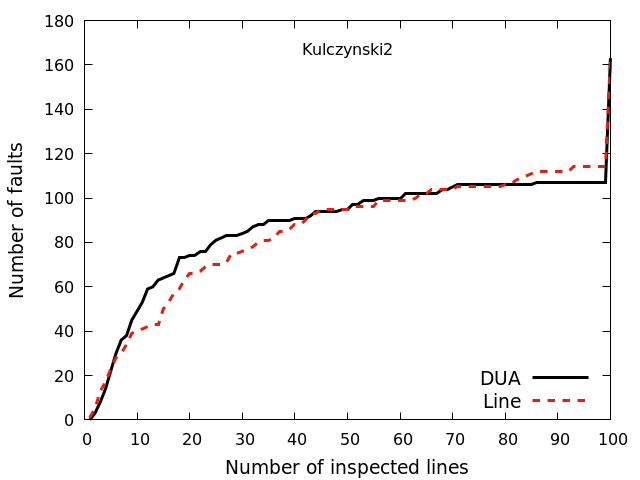}}}}
\mbox{\subfigure[McCon]{\scalebox{.35}{\includegraphics{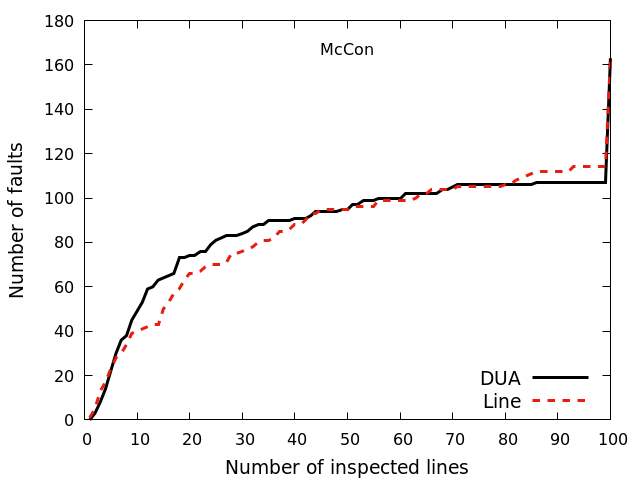}}}\quad
      \subfigure[Jaccard]{\scalebox{.35}{\includegraphics{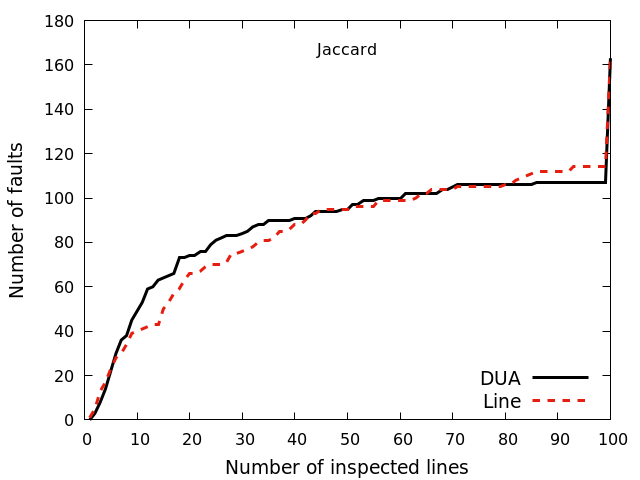}}}} 
\mbox{\subfigure[Zoltar]{\scalebox{.35}{\includegraphics{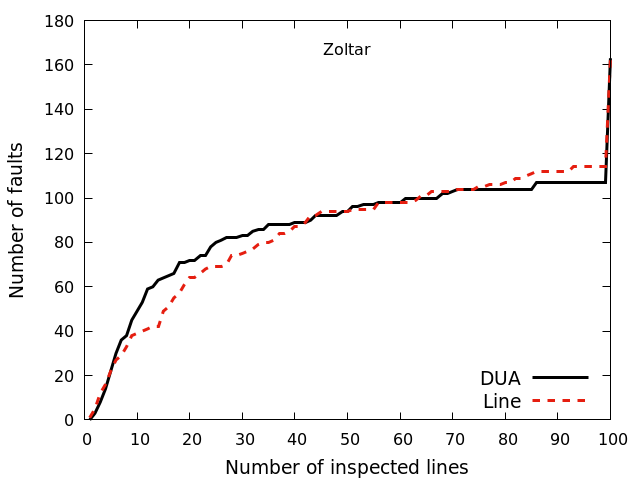}}}\quad
      \subfigure[Tarantula]{\scalebox{.35}{\includegraphics{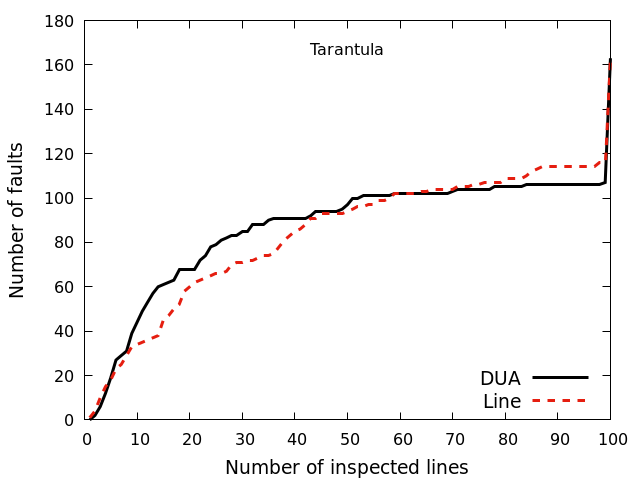}}}}
\caption{Effectiveness of DUA and line spectra in fault localization using different ranking metrics}
\label{fig:effectiveness-1}
\end{figure*}

\begin{figure*}
\centering
\mbox{\subfigure[O$^{p}$]{\scalebox{.35}{\includegraphics{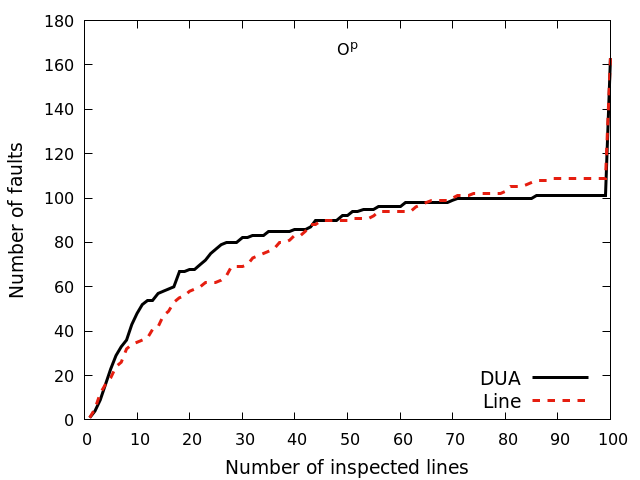}}}\quad
      \subfigure[DRT]{\scalebox{.35}{\includegraphics{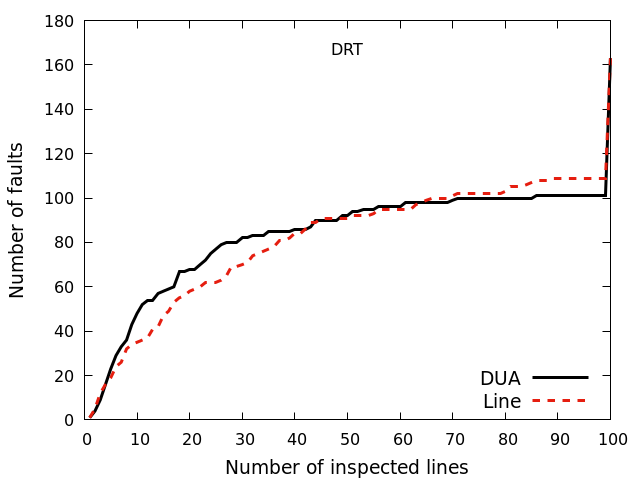}}}}
\mbox{\subfigure[Minus]{\scalebox{.35}{\includegraphics{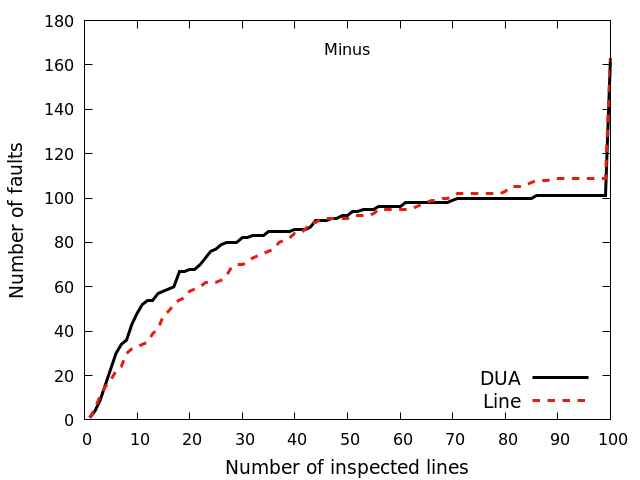}}}\quad
      \subfigure[Wong3]{\scalebox{.35}{\includegraphics{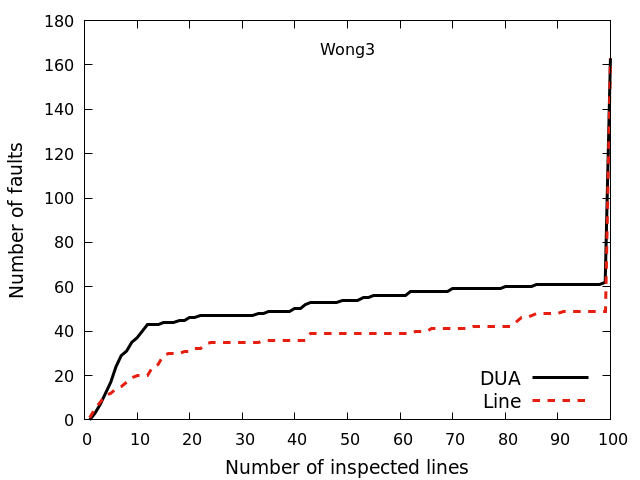}}}}
\caption{Effectiveness of DUA and line spectra in fault localization using different ranking metrics}
\label{fig:effectiveness-2}
\end{figure*}

We compared the effectiveness of data- and control-flow spectra using the ten ranking metrics described in Table~\ref{tb:metrics}. The most suspicious DUAs and lines, ranked according to the metrics, are mapped onto lines to be inspected. Figures~\ref{fig:effectiveness-1} and \ref{fig:effectiveness-2} describe the number of faults located when a developer inspects different amounts of suspicious lines. 

DUA and line spectra have similar behaviors for all ranking metrics, except Wong3. They locate the same number of faults up to 7 investigated lines. After that, and up to 40 lines, DUA spectrum locates more faults. The largest gap is located at 15 lines---for almost all metrics, DUA locates 60 faults whilst line spectrum locates 40 faults. There is no difference between DUA and line spectra for more than 40 lines inspected up to 70 lines. For more than 70 lines, line spectrum performs better.
For Wong3, DUA spectrum performs better after 5 lines, with a constant gap of 20 more faults located. This difference maintains almost constant until the end. However, Wong3 perform poorly locating at most 60 faults when using DUA and 40 with line spectrum. The other metrics locate more than 100 faults inspecting up to 100 lines using line spectrum. DUA spectrum locates about 80 faults by inspecting only 20 lines using Ochiai, Jaccard, Kulczynski2, McCon, $O^p$, and Minus.

Table~\ref{tb:statistics} presents the results of the statistical analysis. The null 
hypothesis is that there is no difference in the effectiveness of data- and control-flow spectra 
to locate faults. The alternative hypothesis is that data-flow spectrum is more effective than 
control-flow spectrum---i.e., data-flow spectrum requires the examination of less suspicious lines 
to locate faults. 


\begin{table*}[ht]
\centering
\caption{Statistical results comparing data- and control-flow}
\label{tb:statistics}
\begin{tabular}{l|r|r|c|rr|rr|rr|rr}
  \hline
  \multirow{2}[2]{*}{Metric} & P-value & \multirow{2}[2]{*}{Effect Size} & \multirow{2}[2]{*}{B-T-W} & \multicolumn{2}{c}{Min} & \multicolumn{2}{|c|}{Q1} & \multicolumn{2}{c|}{Q2} & \multicolumn{2}{c}{Q3} \\
  & \multicolumn{1}{c|}{(\%)} &         & &  DF 	& CF	& DF& CF	&  DF 	& CF &  DF 	& CF\\ 
  \hline
  Ochiai		& 1.59	& 0.032 (I)		& 73-50-40 & 2 & 1 & 9	& 15	& 25	& 39  & 100 & 100 \\
  Kulczynski2 	& 1.66	& 0.023	(I)		& 74-49-40 & 2 & 1 & 9	& 11.5	& 26	& 36  & 100 & 100 \\
  McCon			& 1.66	& 0.023	(I)		& 74-49-40 & 2 & 1 & 9	& 11.5	& 26	& 36  & 100 & 100 \\
  Jaccard		& 1.02	& 0.029	(I)		& 74-49-40 & 2 & 1 & 9	& 11.5	& 26	& 36  & 100 & 100 \\
  Zoltar		& 1.12	& 0.024	(I)		& 75-49-39 & 2 & 1 & 9	& 12.5	& 27	& 37  & 100 & 100 \\
  Tarantula  	& 0.59	& 0.035	(I)		& 75-48-40 & 2 & 1 & 10 & 15	& 27	& 39  & 100 & 100 \\
  {\boldmath$O^p$}	& 1.39	& 0.027	(I)	& 69-57-37 & 1 & 1 & 9	& 13.5	& 30	& 40  & 100 & 100 \\
  DRT			& 2.32	& 0.025	(I)		& 68-57-38 & 1 & 1 & 9	& 13.5	& 30	& 39  & 100 & 100 \\
  Minus			& 2.03	& 0.033	(I)		& 68-57-38 & 1 & 1 & 9	& 14.5	& 30	& 39  & 100 & 100 \\
  Wong3			& 0.01	& 0.331 (S) 	& 46-100-17& 2 & 1 & 12 & 70	& 100	& 100 & 100 & 100 \\
\hline
\end{tabular}
\end{table*}

The \textit{P-value} column presents the results of the significance test. For all metrics, 
DUA spectrum locates more faults inspecting less code with statistical significance (p-value $\leq$ 5\%). 
The \textit{Effect Size} column presents the Cliff's delta results for all metrics, in which 
\textit{I} means that the size is insignificant and \textit{S} is a small effect. Except for Wong3, for which the effect size is small, the other metrics effect size is insignificant.

The \textit{B-T-W} column shows the number of times DUA spectrum was better than (B), tied with (T) or 
worse than (W) line spectrum. It shows that DUA was better than line for near to half of the faults 
(46\% or 75 faults for Ochiai and Zoltar). Line spectrum was better for one-quarter of the bugs 
(e.g., Ochiai, Tarantula, Jaccard). 

Table~\ref{tb:statistics} also presents the minimum number of inspected lines to find a bug (\textit{Min}), the first (\textit{Q1}), second (\textit{Q2}), and third (\textit{Q3}) quartiles for both 
data-flow (DF) and control-flow (CF). DUA spectrum with Ochiai locates half of the faults by inspecting 
25 lines while line spectrum needs 39 lines to reach the same amount of faults. The difference in Q2 varies 
from 14 (Ochiai) to 9 (DRT). The first quartile also indicates that DUA spectrum is more 
effective to locate one-quarter of the faults. It requires the inspection of 9 lines for almost all 
metrics while it can require 15 lines using line spectrum with Ochiai. Both DF and CF were not able 75\% of all faults as shown in the Q3 column.

\subsection{Efficiency of data-flow and control-flow}

To understand the costs of using data- and control-flow spectra, we measured the time spent 
to use both spectra for SFL. The time spent for DF and CF comprises: (1) running all tests, (2) collecting the spectra data, (3) calculating suspiciousness of all program elements, and (4) creating the SFL lists. Table \ref{tab:efficiency} shows the results for the 
efficiency comparison. The \emph{JUnit}, \emph{DF}, and \emph{CF} columns 
show the time spent (in seconds) for executing, respectively, only the JUnit tests, Jaguar 
with data-flow spectra, and Jaguar with control-flow spectra. These values 
consist of the average execution time for all versions of a project for all ten ranking 
metrics.

The \textit{DF/JUnit} column shows the overhead to collect data-flow spectra compared to 
the execution of JUnit. The \textit{CF/JUnit} column presents the overhead to collect 
control-flow spectra compared to JUnit. The average overhead for all programs using data-flow 
and control-flow are 353\% and 102\%, respectively. The \textit{DF/CF} column shows the ratio 
between the data-flow time and control-flow time. It indicates how much extra time is needed 
to execute Jaguar with data-flow compared to control-flow. For JFreeChart, generating a 
list of suspicious DUAs requires 143\% more time than generating a list of suspicious 
statements. However, the total time to generate the list of suspicious DUAs in this case 
is 88 seconds.


\begin{table}[htbp]
\centering
\caption{Data-flow and control-flow efficiency for each program}
\label{tab:efficiency}
\setlength{\tabcolsep}{0.5em}
\begin{tabular}{l|rrr|rrr}
  \hline
  \multirow{2}[2]{*}{Program} & \multicolumn{3}{c|}{Run-time (s)} & \multicolumn{3}{c}{Overhead (\%)} \\
                & \multicolumn{1}{c}{JUnit}     & \multicolumn{1}{c}{DF}    & \multicolumn{1}{c|}{CF}    & DF/JUnit& CF/JUnit & DF/CF \\ 
  \hline
  Commons Lang  & 18.83     & 89.67	& 33.92	& 376.08  & 80.09	& 164.36 \\ 
  Commons Math	& 144.32    & 515.95& 265.16& 257.51  & 83.74	& 94.58 \\ 
  JFreeChart    & 22.27     & 88.62	& 36.36	& 298.0	  & 63.3	& 143.72 \\ 
  Joda-Time	    & 4.88      & 165.42& 47.93	& 3287.71 & 881.67	& 245.1 \\ 
  jsoup         & 4.35      & 22.95	& 11.37	& 427.52  & 161.38	& 101.82 \\ 
  \hline
  Average       & 38.93     & 176.52& 78.95	& 353.43  & 102.8	& 123.58 \\ 
  \hline
\end{tabular}
\end{table}

\section{Discussion}
\label{dis}

To better understand the results of Section~\ref{res}, we discuss and analyze the results next. 
The discussion is organized by our research questions.

\subsection*{Which spectrum is more effective to locate bugs: data- or control-flow?}

The results confirm our hypothesis that data-flow, by providing more detailed information, improves fault localization. DUA spectrum leads the developer to inspect less code than line spectrum with statistical significance for all ranking metrics. 
However, the effect size is insignificant, except for Wong3. DUA spectrum outperforms line spectrum in the range of 7 to 40 investigated lines. After that, there is no difference or line spectrum performs better, which explains the insignificant effect size. For most ranking metrics, control-flow reaches more faults than data-flow after the 70th position. 

Indeed, if the faulty DUA is not the first or second pick, the bug will hardly be located at the top 5 suspicious lines since DUAs are composed of a definition, a use, and possibly a source, if a p-use DUA. Thus, each ranked DUA adds 2 to 3 lines to be investigated. 
Being able to beat line spectrum at the top 10 lines shows that data-flow spectrum singles out specific faulty DUAs to be investigated. Additionally, DUA spectrum demanded less code to be inspected for a large number of faults: 75 out of 163 using Ochiai and Zoltar whilst line spectrum required the inspection of less code only for 40 and 39 faults, respectively, for the same ranking metrics---a difference of 87.5\%.

Around 20 out of 163 (12\%) faults can be located by investigating only the top 5 lines, using either DUA or line spectra. On the other hand, one can reach 40 faults (25\%) by investigating only 9 lines using DUA spectrum and all metrics, excepting Tarantula and Wong3. Furthermore, if the developer investigates up to 25 suspicious lines, s/he will reach half of the faults using Ochiai and DUA spectrum. If the practitioner decides to use line spectrum, s/he will need to inspect 14 more lines to achieve the same result.

Some studies have shown, though, that practitioners may give up on using an SFL technique if the fault is not located at the top 5 picks \cite{parnin2011,kochhar2016}. Since there is no difference between DUA and line spectra up to 7 suspicious lines, the difference is too little to be useful in practice.  
Notwithstanding, sticking to only the first 5 lines is not practical either since too few faults can be located in this short range, in our study, only 12\%. Data-flow spectrum can leverage SFL techniques. However, practitioners should be educated to inspect 10 to 25 lines to reach a quarter to half of the faults, respectively. Other recent results have shown that SFL can help developers to locate faults even when the faulty line is not ranked among the top suspicious ones \cite{souza2018}.

\subsection*{Which spectrum is more efficient to locate bugs: data- or control-flow?}

Table \ref{tab:efficiency} summarized the run-time costs for each project. 
The data-flow suspiciousness lists require from 94\% to 245\% more time to be generated
than the control-flow lists. This result differs from the 38\% overhead presented 
by BA-DUA \cite{araujo2014}. Such a discrepancy is justified by the extra work needed 
to run an SFL technique.

Data-flow implies more information than control-flow: if on the one hand it
provides detailed data for fault localization; on the other hand, more time is needed to 
compute such data. Thus, control-flow is more efficient, providing fault localization 
information with less overhead.

However, the amount of time needed to generate the lists of suspicious DUAs for the medium- 
and large-sized real programs used in this assessment is not prohibitive. For four out of five subjects, 
the time varies from 22 seconds to 2:45 minutes. Only for Commons-Math, the time jumps 
to 8:35 minutes. For this particular program, line spectrum is also expensive to obtain: 4:25 minutes.
Since DUA spectrum can improve the fault localization effectiveness, the extra time to create the 
suspiciousness lists seems affordable at industrial settings.

\section{Threats to validity}
\label{threats}

The internal validity threats to our experiment are related to the Jaguar tool. To mitigate them, 
we manually checked the Jaguar's output using small programs. Regarding the data-flow efficiency results, it is certainly possible to optimize the Jaguar tool to reduce its run-time overhead for both spectra. Our 
chosen strategy aimed to get a first picture of the performance benefits from the recent 
promising data-flow spectra tool (BA-DUA).
As we explain in Section~\ref{exp}, we only selected bugs for which data-flow spectrum was complete. 
Thus, we excluded faults that throws exceptions without handling them and faults in single-block methods. 
Improvements in BA-DUA can allow us to use such faults in future experiments.

All faults were found in their source code repositories. We assumed that each code change between a faulty version and its consecutive fixed version was made to fix the bug. We disregarded changes that do not affect the program behavior such as adding or removing empty lines and line indentation changes. 
For multiple-line faults, we considered the best case scenario. A fault is deemed as found when the top ranked one of its lines is reached. However, this scenario occurs for both spectra. For tie cases, we considered the worst case, in which all lines with the same suspiciousness score of the faulty one are deemed to calculate its position.

Regarding external validity threats, we used five programs from different domains and 
sizes to expose the techniques to different scenarios. Although the programs utilized in 
the experimental assessment are quite heterogeneous, we caution the reader that the 
techniques may present different results for a different set of programs. Also, all programs assessed in this experiment have a single fault per version. The results may differ for programs containing multiple faults.



Regarding construct validity threats, we assessed the techniques using statistical tests and descriptive data. Our experiment was built to evaluate how quickly a technique will reach the fault site. One issue, though, deems the assumption that the ranked elements will be followed as they were presented, which may not actually happen in practice. Another issue is that reaching the bug site does not necessarily mean locating the defect, which is known as perfect bug detection \cite{parnin2011}.

\section{Related work}
\label{related}

Despite the ability to exercise different aspects during a test execution, data-flow 
is rarely used to support SFL. Few initiatives use definition-use 
associations (DUAs) \cite{santelices2009,masri2010} while others use program slicing \cite{lei2012,ju2014} or
different types of data and control dependencies \cite{masri2010,yu2011,eichenger2010}. All these approaches demand
a costly time overhead to collect data-flow spectra.

Santelices et al. \cite{santelices2009} proposed a technique, \textit{avg-SBD}, that combines different 
spectrum types for fault localization: statements, branches, and du-pairs---a variant 
of data-flow information that takes into account only c-use DUAs. They also introduce an 
approximate DUA spectrum, \textit{avg-SBD approx}, which has lower overhead at run-time. 
The results, assessed using small- to medium-sized programs, showed that their approach 
locates more faults than each spectrum isolated. 
They also showed that different faults are best located by different coverage types. 

Masri \cite{masri2010} uses a spectrum called \textit{Dynamic Information Flow 
Analysis} (DIFA) for fault localization. DIFA is composed of interactions 
performed in an execution, including inter-procedural data- and control-flow 
dependences from statements and variables. The author compared DIFA with statement, branch, 
and DUA spectra using 18 faults of medium-sized programs. DIFA presented better 
performance to locate the bugs. However, the time to generate DIFA spectra 
was 2--10 times slower than DUA and branch spectra.

Hemmati \cite{hemmati2015} assessed the ability of data- and control-flow coverage criteria 
to reveal faults on programs from the Defects4J database. Du-pair criterion covered 79\% of faults that 
were not covered by control-flow criteria. He also showed that 15\% of all faults were not 
covered by any criterion. The author did not report issues regarding run-time costs.

Yu et al. \cite{yu2011} presented the \textit{LOUPE} technique, which combines data- and 
control-dependencies for fault localization. The output is a list of statements. 
Their experiments using the Siemens suite presented better performance than the compared 
techniques. They also assessed LOUPE in the space program, which took 100 minutes to execute.
Eichinger et al. \cite{eichenger2010} propose the data-flow enabled call graphs (DEC graphs), which is a method 
call graph with data-flow information. DEC graphs register also parameters and method-return 
values. To reduce the number of possible parameter values, they discretize numerical parameter 
and return values using Data Mining approaches. They assessed the technique including ten faults in 
the Weka program---which has 301 KLOC. The authors do not reported the run-time overhead of the technique.

As with the works of Santelices et al. \cite{santelices2009} and Masri \cite{masri2010}, our study compares 
DUA and statement spectra to investigate their effectiveness for fault localization. 
Differently from the previous works, we assessed both spectra using ten different ranking 
metrics. Moreover, we used medium- to large-sized programs containing real faults. 
We also evaluated their performance to locate faults for absolute values and compared the overhead of SFL techniques based on data- and control-flow spectra. Finally, we carried out a statistical analysis in our assessment, 
showing that data-flow spectrum can be useful for fault localization at an affordable cost. 

\section{Conclusions}
\label{concl}

We carried out an experiment comparing the use of data-flow spectrum (definition-use 
associations---DUAs) and control-flow spectrum (lines) in fault localization. Our
goal was to corroborate the body of evidence with more robust results, showing that data-flow is more
effective than control-flow in fault localization. In addition, we aimed
to verify whether new approaches to collect DUAs would allow the application
of data-flow-based SFL in programs similar to those developed in the industry.
We used 163 real faulty versions of five industry-like programs whose sizes vary from 10 to 96 KLOC. 

DUA spectrum located more bugs inspecting less code for all ranking metrics with statistical significance. DUA spectrum outperforms line spectrum in the range from 7 to 40 inspected lines. By inspecting from 10 to 25 lines, a practitioner will reach the bug site for a quarter to half of the faults. Line spectrum requires to investigate 14 more lines to reach half of the faults. After 70 inspected lines, line spectrum can reach more bugs than DUA spectrum. As each DUA has 2 or 3 lines, its performance is overcome by line spectrum as the number of inspected lines increases.

It is known that practitioners expect the fault to be located in the first 5 most suspicious lines. Nevertheless, such a tight target is successful for only 12\% of the faults in our experiment. SFL will hardly be adopted in practice with this small success rate. Data-flow spectrum can help the adoption of SFL techniques at industrial settings, but the practitioners should be educated to inspect a few more code excerpts to take advantage of SFL techniques. 

Our results suggest that from 10 to 25 lines would significantly improve the number of faults hit by SFL techniques using data-flow spectrum. Regarding efficiency, the overall time for the subjects of our experiment varies from 22 seconds to 8:35 minutes using data-flow spectrum, which indicates that its use is feasible for many programs developed by the industry.
Thus, the hypothesis that data-flow generates more detailed information that improves fault localization was verified in our experiment. Although more time is needed to collect such an information, the overhead is affordable for real-world software.

In future work, we will exploit contextual information associated with data-flow spectrum, 
especially the use of variables from the most suspicious DUAs to locate faults. 
We will also carry out user studies to evaluate whether information regarding suspicious variables improves fault localization.


\section*{Acknowledgment}

This research was supported by FAPESP (S\~ao Paulo Research Foundation), processes 13/24992-2, 14/23030-5, and 14/50937-1; by CNPq process 311174/2017-5; and by the Research Support Center on Open Source Software (NAPSoL) of the University of S\~ao Paulo. 

\bibliographystyle{IEEEtran}
\bibliography{df-cf-sfl} 
\end{document}